\documentclass[12pt]{iopart}
\usepackage{graphicx}

\begin{document}

\title[False discovery rate]{False discovery rate: setting the probability of false claim of detection}

\author{L~Baggio\footnote[2]{ presently at Institute for Cosmic Ray Research (Univ. of Tokyo)
5-1-5 Kashiwanoha, Kashiwa-shi, Chiba-ken, 277-8582, Japan}
and G~A~Prodi
}

\address{Dipartimento di Fisica, Universit\`a di Trento and INFN, Gruppo collegato di Trento, Sezione di Padova, via Sommarive, 14, 38050, Povo, TN, Italy}

\ead{baggio@science.unitn.it}

\begin{abstract}
When testing multiple hypothesis in a survey --e.g. many different source locations, template waveforms, and so on-- the final result consists in a set of confidence intervals, each one at a desired confidence level. But the probability that at least one of these intervals does not cover the true value increases with the number of trials. With a sufficiently large array of confidence intervals, one can be sure that at least one is missing the true value. In particular, the probability of false claim of detection becomes not negligible. In order to compensate for this, one should increase the confidence level, at the price of a reduced detection power.
False discovery rate control\cite{Benjamini95} is a relatively new statistical procedure that bounds the number of mistakes made when performing multiple hypothesis tests. We shall review this method, discussing exercise applications to the field of gravitational wave surveys.

\end{abstract}


\submitto{\CQG}


\section{Introduction}

The motivation for controlling the false discovery rate (FDR) --i.e. the fraction of false alarms in a collection of candidate detections-- jumped to our attention as we were involved in data analysis for the IGEC~\cite{IGEC-PRD}, the network of resonant detectors that searched for coincident burst gravitational wave (GW) signals in the years 1997-2000. Even if the detectors involved in IGEC were rather similar, there were obvious configurations (special choice of detector pairs, three-fold instead of double coincidence) or cuts of the data (higher or lower threshold on event amplitude) characterized by lower background counts, or higher duty time. We did not have \textit{a priori} a good reason to prefer one configuration or cut more than others, as we do not know a priori the intensity of the signal, hence the efficiency. Therefore, we decided at the beginning a fairly long list of interesting choices, in order to perform many analyses in parallel, and eventually to quote the results for each trial.

The results were expressed as confidence intervals on the expectation value for the number of counts in coincidence due to GW.
When unveiling the final results, one of the confidence intervals at 90\% coverage was not including the null hypothesis (i.e. zero counts). Of course this can be somewhat expected by chance when the number of trials is very high. It was possible to compute accurately that with 30\% probability we had a chance that at least one of the tests falsely rejected the null hypothesis.

\begin{table}[t]
\caption{\label{tab:notation} Quick-reference notation chart for the variables used in \sref{sec:description}. $m$ is the total number of performed tests (trial factor), $m_0$ and $m_1$ the real number of underlying off-source and on-source tests. The number of \textit{actually} positive tests is R, given by S true positives and B spurious claims. An ideal experiment would neither treat background as signal (type I error) nor do the reverse (type II errors).}
\centering
\begin{indented}
\item[]
\begin{tabular}[]{c|c|c|c|}
\br
 & \parbox[h][][c]{2.6cm}{\centering Null Retained\\\textit{(cannot reject)}} &  \parbox[h][][c]{4cm}{\centering Reject Null\\\textit{(i.e. accept alternative)}} & Total\\
\hline
\parbox[h][][c]{2.6cm}{Null ($H_0$) True\\(\textit{background})}& \textbf{$m_0$-B} & \parbox[h][1cm][c]{4cm}{\centering \textbf{B}\\ \textit{Type I Error}} & $m_0$\\
\hline
\parbox[h][][c]{2.6cm}{Alternative True\\(\textit{signal})} & \parbox[h][1cm][c]{3cm}{\centering \textbf{$\beta=m_1$-S}\\\textit{Type II Error}} & \parbox[h][][c]{5cm}{\centering \textbf{S} \\ \textit{Detected signals}}& $m_1$\\
\hline
&  \textbf{$m$-R} & \parbox[h][1cm][c]{5cm}{\centering \textbf{R=B+S}\\\textit{Reported signal candidates}} & \textbf{$m$}\\
\hline
\end{tabular}
\end{indented}
\end{table}

The probability of at least one false claim in a set of trials is known as \textit{family-wise error rate} (FWER). It is not difficult to devise a method to control this quantity \textit{before} going to the results: we just have to increase the confidence in the single trial (say 99\%, or 99.99\% coverage) in order to keep the FWER much lower than one. The drawback is that the resulting confidence interval would be much larger, and consequently the power of the search would fall dramatically. This is a consequence of the request that \textit{not even in a single case} the null hypothesis is rejected when it is true.

A very reasonable compromise was suggested by Benjamini and Hochberg\cite{Benjamini95}. They remark that in many practical cases, when having one or more false claim is not by itself unacceptable, we could just be happy if --on average-- \textit{most} of the claims were real. In other words, they propose to bound FDR instead of FWER.

There are many topics in GW search which would benefit from this kind of procedure. For instance:

\begin{itemize}
\item all sky surveys: many source directions and polarizations are tried in parallel;
\item template banks;
\item eyes-wide-open searches: many alternative analysis pipelines, with different amplitude thresholds, signal duration, and so on are applied on the same data.
\item periodic updates of results: every new science run is a chance for a ``discovery" (``Maybe next one is the good one");
\item Many graphical representations or aggregations of the data (``With a slight change in the binning, the `signal' shows up better")
\end{itemize}

This work means not to be a complete review of the state-of-art techniques about FDR control, but hopefully it will be a stimulus for whoever is involved in multiple-test data analysis issues.

In the following sections, we shall use the notation reported in \tref{tab:notation}.

\section{Description of the method}
\label{sec:description}

\subsection{Preliminary remarks}

In order to decide whether the results of a measurement are compatible with being generated by noise only (\textit{null hypothesis}, H0) or instead they contain a signal (\textit{alternative hypothesis}, H1) the textbook procedure is to set up a test statistic $t$ from the measures themselves. If $F_0(t)$ is the distribution of $t$ when the H0 holds, then the $p$-value of $t$ is defined as $p = F_0(t) = \Pr(t_0>t|\forall t_0)$. By construction $p$ is uniformly distributed between 0 and 1:

\begin{equation}
\label{eq:background}
\Pr(p<p_0|0 \le p_0 \le 1)=p_0
\end{equation}

It is of paramount importance that the distribution $F_0$ is known. It is always wise to check a priori models with a goodness-of-fit test, when there are enough off-source data available. This is not always the case, but often there are surrogate procedures (e.g. data permutation) which give fresh independent samples of the background process, removing at the same time the effect of real signals, if any are present in the data. For instance, in the case of IGEC, the \textit{resampling procedure} consisted in adding a delay to the time reference of one of the detectors in the network, such that the coincident signal is lost, while the background expectation value of coincidence counts is approximately unchanged. In case the data are not compliant with the model, at worse resampled data may allow to estimate $F_0$ by empirical fit.

As for H1, it is usually unknown, but for our purposes it is sufficient assuming that the signal can be distinguished from the noise, i.e. $\Pr(p<p_0|0 \le p_0 \le 1) \ne p_0$. The sketch in \fref{fig:FDR} (\textit{top left}) illustrates the concept.

For a single hypothesis test, the condition ``reject null if $p<\alpha$" leads to false positives with probability $\alpha$.
In case of multiple tests, we deal with a set ${\bf p} \equiv \{p_1, p_2, \dots p_m\}$ of $p$-levels, which need not to derive from the same test statistics, nor they should refer to same tested null hypothesis. $m$ is called the trial factor.
We select discoveries using a threshold $T({\bf p})$: ``reject null if $p_j<T({\bf p})$".

\subsection{Controlling Type I errors (B)}

The \textit{uncorrected testing} would just use the same threshold for each test: $T({\bf p})= \alpha$. The probability that at least one rejection is wrong grows as
$P(B>0) = 1- (1- \alpha)^m \approx m\alpha$.

Therefore, as in the IGEC case, false discovery is guaranteed for $m$ large enough.

The other extreme solution, usually referred to as the \textit{Bonferroni procedure} \cite{Bonferroni}, controls the FWER in the most stringent manner, by requiring that $P(B>0) \le \alpha$. This is achieved by the choice $T({\bf p})= \alpha/m$
While this approach makes mistakes rare, the cost is low efficiency ($S\approx0$).

\subsection{Controlling false discovery rate (B/R)}
\label{ssec:fdr}

In order to trade-off between B=0 and S=0, the FDR control focuses on the ratio of false discoveries to the total number of claims:

\begin{equation}
\label{eq:fdr}
FDR \equiv \left\{ {\begin{array}{*{20}c}
   {{B \mathord{\left/
 {\vphantom {B R}} \right.
 \kern-\nulldelimiterspace} R} \equiv {B \mathord{\left/
 {\vphantom {B {(B + S)}}} \right.
 \kern-\nulldelimiterspace} {(B + S)}}} & {{\rm if  }R > 0}  \\
   0 & {{\rm if  }R = 0}  \\
\end{array}} \right.
\end{equation}

\begin{figure}[t]
\centering
\includegraphics[width=8cm]{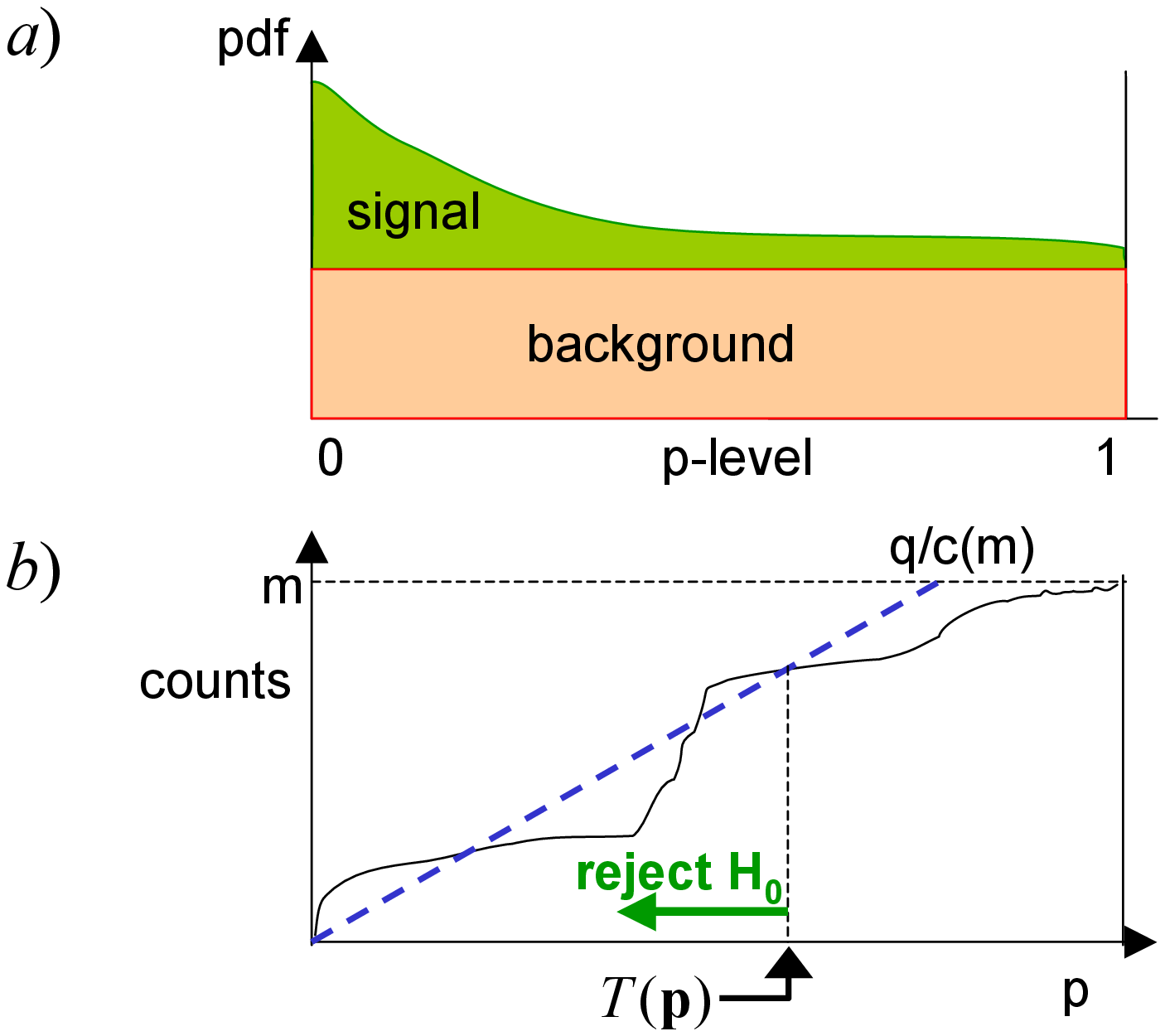}
\includegraphics[width=7cm]{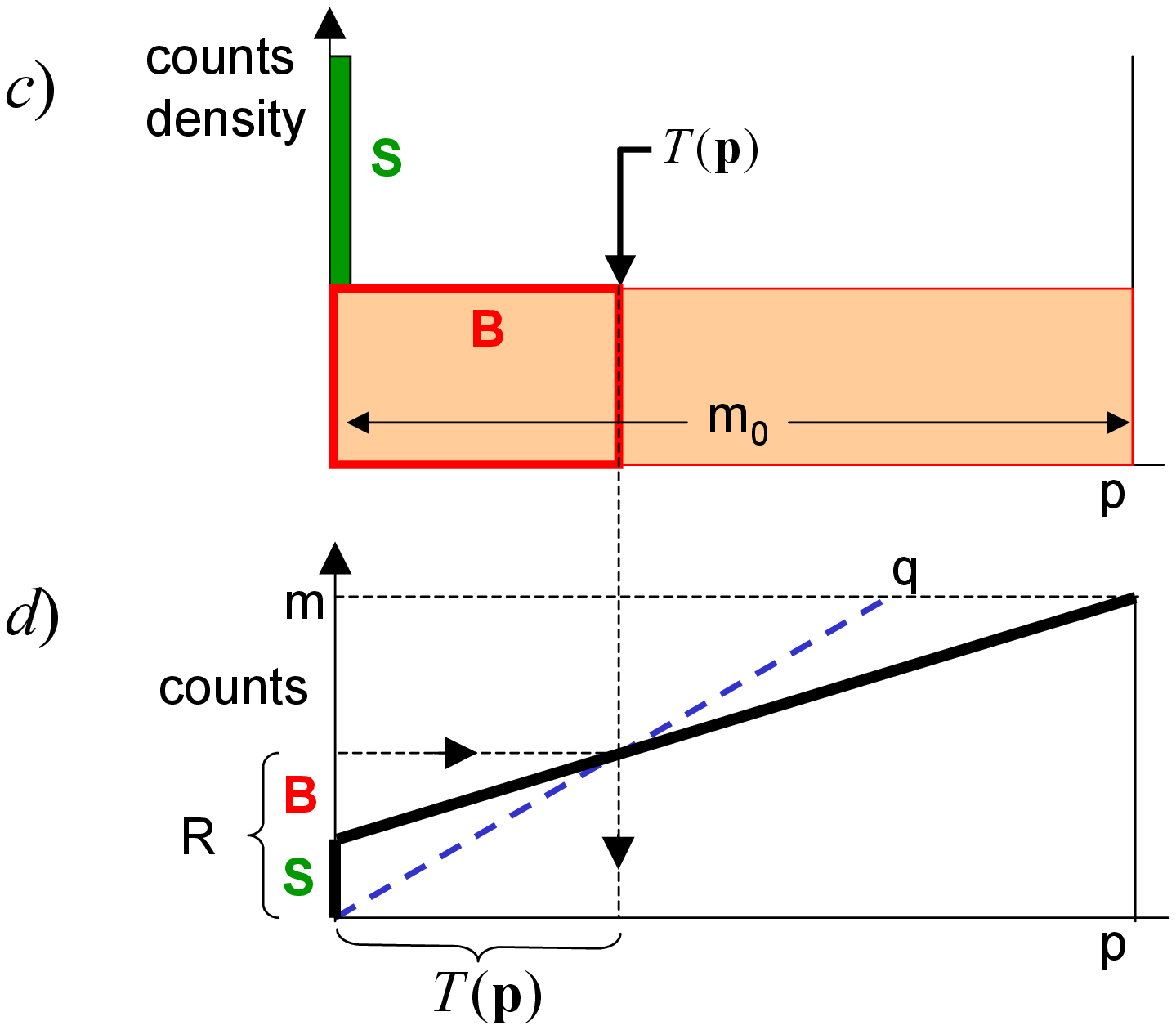}
\caption{\label{fig:FDR}(\textit{a}) The probability density function of $p$-values when data come from a mixed model can be thought as the sum of a uniform distribution (background) and a biased one (signal). (\textit{b}) The Benjamini-Hochberg procedure (BH) consists in plotting the cumulative histogram of the $p$-values of the $m$ trials (\textit{continuos line}) and looking for intersections with a line drawn from the origin and with slope equal to $m \cdot c(m)/q$ (\textit{dashed line}). The null hypothesis is rejected for all data with $p$-value between 0 and the abscissa of the highest intersection point.\\(\textit{c}) Sketch of histogram of $p$-values and (\textit{d}) corresponding cumulative histogram, in a case of easily separable signals. The BH procedure applied to this case can be easily shown to control FDR (see \sref{ssec:fdr} for details).}
\end{figure}

This can be done with a proper choice of $T({\bf p})$. The original procedure suggested by Benjamini and Hochberg (BH) is extremely simple, involving only trivial algebraic operations. It consists in the following steps.
\begin{itemize}
\item sort the $p$-values in ascending order: $\{p_1, p_2, \dots p_m | i < j \Rightarrow p_i \le p_j \}$;
\item choose your desired FDR $q$ (in case no signal source is actually present during the observation, then the procedure is equivalent to the Bonferroni procedure with $\alpha=q$);
\item define $c(m)=1$ if $p$-values are independent or positively correlated; otherwise
$c(m) = \sum\nolimits_{j = 1}^m {{1 \mathord{\left/ {\vphantom {1 j}} \right. \kern-\nulldelimiterspace} j}} $;
\item determine the threshold $T({\bf p})=p_j$ by finding the index $j$ such that $p_k>k(q/m) /c(m)$ when $k>j$ (see \fref{fig:FDR} for a visual representation of this condition).
\end{itemize}

The above procedure with $c(m)=1$ was shown \cite{Benjamini95} to control the expectation value\footnote{Of course, the quantity FDR is a random variable, as well as the $p$-values.} of FDR at least at level $q$ in the case when all $m$ tests are independent. However, it was later proved to control FDR when tests are positively correlated \cite{Benjamini01} (for instance, multivariate normal data where the covariance matrix has all positive elements). The alternative definition of $c(m)$ given above controls FDR in the most general case \cite{Benjamini01}, but at the cost of reduced efficiency.

There is a nice back-of-the-envelope plausibility argument which can be found in ~\cite{Miller01} for the simple case when signals are easily separable (e.g. signals with high signal-to-noise ratios). In this case we expect their $p$-level to be very low, and correspondingly in the cumulative histogram of $p$-levels we shall see a step with height $S$ near $p \approx 0$, see \fref{fig:FDR} (\textit{bottom right}). We see also that there is only one intersection point for the BH procedure, such that
\begin{equation}
\label{eq:miller}
T({\bf p})/R = q/m
\end{equation}
On the other hand, the threshold $T({\bf p})$ can be expressed on average by $B/m_0$ (this is a special case of \eref{eq:background}). Substituting this value in \eref{eq:miller} we obtain
\begin{equation}
B/R = q m_0/m \le q
\end{equation}
For a rigorous proof see \cite{Benjamini01}.

\section{Numerical test of the method}
\label{sec:example}


We now demonstrate this procedure with a simple example. Suppose we are given the results of 50 counting experiments, labeled by the index $i$. Their background is modeled as a Poisson random variable, with the same\footnote{To avoid degeneracy due to the discreteness of the test statistic (many results collapsing at the same $p$-values), we actually spread the background of the experiments in a range $\pm1\%$ around $N_b$).} known expectation value $N_b$ for all $i$.

We consider two possible cases: in the first one, we draw 50 independent measures, in the other case we generate correlation by summing neighbor bins (i.e., if $n_c^i$ represent independent counts in the $i$-th bin, then the 50 correlated counts ${n'}_c^i$  are defined as ${n'}_c^i=n_c^i+n_c^{i-1}$, where $n_c^{0}\equiv n_c^{50}$).
We investigated different background levels (from $N_b=0.01$ to $N_b=50$) and different number of detected signals ($N_s=0-6$), assuming --for sake of simplicity-- that each bin can have either one or zero counts due to true signals.

In order to decide the presence of a signal we use the one-tail Poisson probability for the expected number of counts in each bin. In \tref{tab:example1} the results of a Monte Carlo simulation are shown. For each configuration (differing by average background and extent of true signals) we compute the average number of claims $R$, i.e. the number of bins for which the null hypothesis is rejected. We present the results for Bonferroni and BH tests, both tuned to bound the FWER at 1\% when no signal is present.

Both procedures are working as expected, controlling the FWER and the FDR respectively at the desired level. For high background values they give as expected similar results. On the other side the efficiency of the Bonferroni procedure falls to zero for $N_b>0.01$, while the BH procedure is still effective, up to $N_b=0.05$ in this example.

\begin{figure}[t]
\centering
\includegraphics[width=10.9cm]{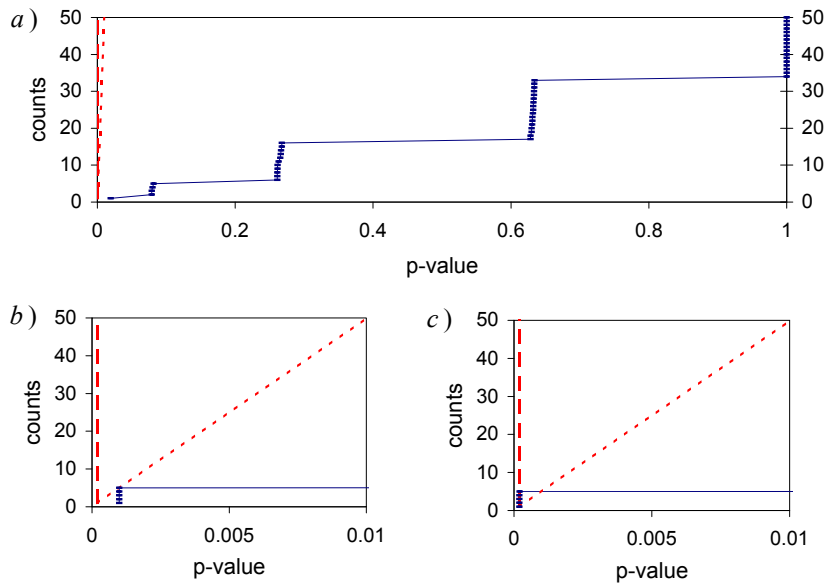}
\caption{\label{fig:example}A few samples from the Monte Carlo used to produce \tref{tab:example1} are displayed in detail. They refer to $N_s=5$, and the background is (\textit{a}) $N_b=50$ (\textit{b}) $N_b=0.5$  (\textit{c}) $N_b=0.01$. In the plots above the cumulative histogram of the $p$-values is compared with the threshold given by the Bonferroni (\broken) and the BH (\dashed) procedures.}
\end{figure}

In \fref{fig:example} we can visualize how the BH procedure manage to grasp the signals promptly, as the background level lowers (see also \fref{fig:FDR}).


\begin{table}[t]
\noindent\caption{\label{tab:example1} Results of the simulation described in \sref{sec:example}. The first column lists values of $N_s$, the other columns refer to differ values of $N_b$, as listed in the first row. Each entry corresponding to a $\{N_s,N_b\}$ couple is composed by values, the upper one refers to the Bonferroni procedure, the lower to the BH procedure. These values are averaged over 40000 samples and the statistical precision is of the order of 0.005.}
\begin{small}
\centering
\lineup
\begin{tabular}[]{ccccccccccccc}
\br
$N_s$ 	& \textbf{ 0.01 }	& \textbf{ 0.02 }	& \textbf{ 0.05 }	& \textbf{ 0.1 }	& \textbf{ 0.2 }	& \textbf{ 0.5 }	& \textbf{ 1 }	& \textbf{ 5 }	& \textbf{ 10 }	& \textbf{ 50 }	\\
\hline \textbf{ 0 }	& 0.005	&-	&-	&  $10^{-4}$	& 3$\cdot10^{-4}$	& 0.003	& 0.007	& 0.008	& 0.003	& 0.004	\\
	& 0.005	& 2$\cdot10^{-4}$	&-	& $10^{-4}$	& 3$\cdot10^{-4}$	& 0.003	& 0.007	& 0.008	& 0.003	& 0.004	\\
\hline \textbf{ 1 }	& 1.005	& 4$\cdot10^{-4}$	& 0.001	& 0.002	& 0.005	& 0.013	& 0.028	& 0.012	& 0.004	& 0.005	\\
	& 1.010	& 0.019	& 0.001	& 0.002	& 0.005	& 0.013	& 0.028	& 0.012	& 0.004	& 0.005	\\
\hline \textbf{ 2 }	& 2.004	& 5$\cdot10^{-4}$	& 0.002	& 0.004	& 0.009	& 0.021	& 0.047	& 0.016	& 0.005	& 0.005	\\
	& 2.010	& 2.019	& 0.002	& 0.004	& 0.009	& 0.021	& 0.047	& 0.016	& 0.005	& 0.005	\\
\hline \textbf{ 3 }	& 3.005	& 0.001	& 0.003	& 0.006	& 0.013	& 0.032	& 0.069	& 0.021	& 0.006	& 0.005	\\
	& 3.010	& 3.019	& 0.004	& 0.006	& 0.013	& 0.032	& 0.069	& 0.021	& 0.006	& 0.005	\\
\hline \textbf{ 4 }	& 4.005	& 0.001	& 0.004	& 0.008	& 0.017	& 0.043	& 0.086	& 0.027	& 0.007	& 0.006	\\
	& 4.010	& 4.018	& 0.125	& 0.008	& 0.017	& 0.043	& 0.086	& 0.027	& 0.007	& 0.006	\\
\hline \textbf{ 5 }	& 5.004	& 0.002	& 0.005	& 0.009	& 0.020	& 0.053	& 0.106	& 0.029	& 0.008	& 0.006	\\
	& 5.009	& 5.018	& 5.046	& 0.009	& 0.020	& 0.053	& 0.106	& 0.029	& 0.008	& 0.006	\\
\hline \textbf{ 6 }	& 6.004	& 0.002	& 0.006	& 0.013	& 0.024	& 0.061	& 0.124	& 0.034	& 0.010	& 0.007	\\
	& 6.009	& 6.017	& 6.043	& 0.013	& 0.024	& 0.061	& 0.124	& 0.034	& 0.010	& 0.008	\\
\br
\end{tabular}
\end{small}
\end{table}

\begin{table}[t]
\noindent{\caption{\label{tab:example2} Same as \tref{tab:example1} but for the case of correlated measures.}}
\begin{small}
\centering
\lineup
\begin{tabular}[]{ccccccccccccc}
\br
$N_s$ 		& \textbf{ 0.01 }	& \textbf{ 0.02 }	& \textbf{ 0.05 }	& \textbf{ 0.1 }	& \textbf{ 0.2 }	& \textbf{ 0.5 }	& \textbf{ 1 }	& \textbf{ 5 }	& \textbf{ 10 }	& \textbf{ 50 }	\\
\hline \textbf{ 0 }	& 0.006	&-	&-	& $10^{-4}$	& 2$\cdot10^{-4}$	& 0.003	& 0.008	& 0.008	& 0.003	& 0.004	\\
	& 0.010	& 0.010	&-	& $10^{-4}$	& 2$\cdot10^{-4}$	& 0.003	& 0.009	& 0.008	& 0.003	& 0.004	\\
\hline \textbf{ 1 }	& 1.005	& 3$\cdot10^{-4}$	& 0.001	& 0.002	& 0.005	& 0.013	& 0.030	& 0.012	& 0.004	& 0.005	\\
	& 1.010	& 0.029	& 0.001	& 0.002	& 0.005	& 0.013	& 0.031	& 0.012	& 0.004	& 0.005	\\
\hline \textbf{ 2 }	& 2.005	& 7$\cdot10^{-4}$	& 0.002	& 0.004	& 0.008	& 0.023	& 0.046	& 0.017	& 0.005	& 0.005	\\
	& 2.009	& 2.018	& 0.006	& 0.004	& 0.008	& 0.023	& 0.047	& 0.017	& 0.005	& 0.006	\\
\hline \textbf{ 3 }	& 3.004	& 0.001	& 0.003	& 0.005	& 0.012	& 0.032	& 0.067	& 0.022	& 0.006	& 0.005	\\
	& 3.009	& 3.019	& 0.060	& 0.005	& 0.012	& 0.032	& 0.068	& 0.022	& 0.006	& 0.006	\\
\hline \textbf{ 4 }	& 4.005	& 0.002	& 0.004	& 0.008	& 0.017	& 0.043	& 0.084	& 0.025	& 0.007	& 0.005	\\
	& 4.009	& 4.017	& 0.143	& 0.008	& 0.017	& 0.043	& 0.085	& 0.025	& 0.007	& 0.006	\\
\hline \textbf{ 5 }	& 5.004	& 0.002	& 0.005	& 0.010	& 0.019	& 0.051	& 0.107	& 0.029	& 0.008	& 0.007	\\
	& 5.009	& 5.018	& 5.047	& 0.010	& 0.019	& 0.051	& 0.108	& 0.029	& 0.008	& 0.007	\\
\hline \textbf{ 6 }	& 6.004	& 0.003	& 0.007	& 0.011	& 0.024	& 0.061	& 0.127	& 0.035	& 0.010	& 0.007	\\
	& 6.009	& 6.016	& 6.044	& 0.013	& 0.024	& 0.061	& 0.127	& 0.035	& 0.010	& 0.007	\\

\br
\end{tabular}
\end{small}
\end{table}

\section{Conclusions}

When multiple tests are tried for the same data set, controlling FDR seems in general a wiser idea than just limiting type-I errors. Robust but simple procedures exist which (conservatively) control FDR in positively correlated tests, and also in the more general case (but at the cost of reduced efficiency).

This idea is relatively new in the astrophysics community, and we are not aware of any application in the GW community. Its application should be encouraged. Notice however that BH procedure is not the only one, and more complex --but approximate-- strategies have been investigated (see for instance \cite{Storey02, Yekutieli99}).

\section{Acknowledgments}

We are indebted to James T. Linnemann (MSU) for introducing us to FDR.

Baggio acknowledges the hospitality of ICRR and the grant of Tokyo University.

\end{document}